\documentclass[twocolumn,natbib]{aastex61}

\received{ }
\revised{  }
\accepted{ }
\submitjournal{ApJ}

\shorttitle{The Clarke exobelt}
\shortauthors{Socas-Navarro}
\newcommand{\degree}{$^{\mathrm{o}}$}
\newcommand{\chio}{$\chi_{\mathrm{o}}$}
\newcommand{\chimax}{$\chi_{\mathrm{max}}$}

\begin{document}

\title{Possible Photometric Signatures of Moderately Advanced
  Civilizations: \\
  The Clarke Exobelt}

\correspondingauthor{Hector Socas-Navarro}
\email{hsocas@iac.es}

\author[0000-0001-9896-4622]{Hector Socas-Navarro}
\affiliation{Instituto de Astrof\'\i sica de Canarias, Avda V\'\i a L\' actea S/N, La Laguna, 38205, Tenerife, Spain}
\affiliation{Departamento de Astrof\'\i sica, Universidad de La Laguna, La Laguna, 38205, Tenerife, Spain}



\begin{abstract}

This paper puts forward a possible new indicator for the presence of
moderately advanced civilizations on transiting exoplanets. The idea
is to examine the region of space around a planet where potential
geostationary or geosynchronous satellites would orbit (herafter, the
Clarke exobelt). Civilizations with a high density of devices and/or
space junk in that region, but otherwise similar to ours in terms of
space technology (our working definition of ``moderately advanced''),
may leave a noticeable imprint on the light curve of the parent
star. The main contribution to such signature comes from the exobelt
edge, where its opacity is maximum due to geometrical
projection. Numerical simulations have been conducted for a
variety of possible scenarios. In some cases, a Clarke exobelt with a
fractional face-on opacity of $\sim$10$^{-4}$ would be easily
observable with existing instrumentation. Simulations of Clarke
exobelts and natural rings are used to quantify how they can be
distinguished by their light curve.

\end{abstract}

%keywords = {astrobiology, extraterrestrial intelligence, intergalactic medium, radio continuum: general},



\section{Introduction}
\label{sec:intro}

The discovery of thousands of exoplanets in recent years has sparked a
surge of research on potential atmospheric biomarkers (see e.g., the
recent reviews by \citealt{SKP+17}; \citealt{G17} and references
therein). Future giant telescopes, such as the JWST, GMT, E-ELT or
TMT, are expected to provide detailed analyses of the atmospheric
composition of transiting exoplanets (\citealt{H16}; \citealt{SDV09};
\citealt{ACH+6}; \citealt{Q15}; \citealt{WLM+14}).  For the first
time in history, humankind would be in a position to search for direct
unequivocal evidence of life on planets around other stars.

While searching for exolife is starting to appear as a feasible
endeavor, the search for intelligent life, or even technological
civilizations, seems almost completely hopeless except under very
specific circumstances. The main problem is the lack of suitable
``technomarkers'' (also referred to as technosignatures,
\citealt{SLF+10}), indicators that, analogously to the biomarkers,
would unequivocally reveal the presence of technology.

Existing proposals in the literature for observable technomarkers are
of extremely speculative nature. \citet{LL17a} analyze the possibility
that the enigmatic fast radio bursts might be the propulsion system of
an advanced interstellar form of transportation. \citet{FE11} put
forward that depletion of certain metallic elements in a stellar
debris disk could be a sign of extensive asteroid mining by an
advanced space-faring civilization. \citet{H02} used data from the
Compton Gamma Ray Observatory to seek traces of antimatter used as an
alien power source. \cite{KSLG15} simulate the observational signature
of giant mirrors put into orbit to illuminate the dark side of a
planet. However, the indicator that has attracted the most attention
and has been more actively pursued is the concept of Dyson spheres
(\citealt{D60}), consisting of astroengineering-scale artificial
structures hypothetically employed to harvest power from the parent
star. These spheres and other observationally equivalent
megastructures would produce occultations of stellar light and excess
infrared emission. The most extensive systematic search for this
indicator is the \^{G} survey (see \citealt{WMS+14}; \citealt{WGS+14};
\citealt{GWM+15}; \citealt{WCZ+17}).

The main problem with all of the indicators mentioned above is that
they would be produced by civilizations with extremely advanced
technologies and all of the underlying uncertainties in this uncharted
territory increase significantly as we move further away from our own
experience. Philosophical and technical questions like what are their
motivations, would anyone care to build a megastructure if they had
already developed nuclear fusion or do such advanced species even
exist, become more frequent and difficult to answer. For that reason,
it is of great interest to develop technomarkers closer to our own
technological level. Radio emissions constitute the only such
indicator that has been actively explored thus far. However, emissions
comparable to ours would be virtually undetectable against the
background at interstellar distances unless they were specifically
targeted in our direction. In fact, all SETI (Search for
Extraterrestrial Intelligence) surveys for radio signals have turned
out empty. The current state of non-detection should not be viewed as
evidence of ETI absence but as a measure of the enormous difficulties
involved in the task (\citealt{SETI09}). Rather than discouraged,
radio SETI efforts have been augmented with other currently ongoing
projects (Breakthrough Listen, Allen Telescope Array and SERENDIP, see
e.g. \citealt{ESF+17}; \citealt{HRT+16}; \citealt{CMC+17} and
references therein), in line with the philosophy that a small effort
is worthwhile, given the tremendous implications of a potential
success.

This paper examines a novel technomarker, the Clarke exobelt
(hereafter CEB), for hypothetical civilizations at our current level
of technological development (at least in terms of space engineering)
but making a heavier use of their planetary space environment. We
shall refer to this as a ``moderately advanced civilization'', to
distinguish from the much more advanced engineering capabilities
required for the other indicators mentioned above. The CEB is formed
by all objects, including functioning devices and space junk, in
geostationary and geosynchronous orbits around a planet. The following
sections show the results of numerical simulations demonstrating that,
under certain not too implausible circumstances, a CEB would be
detectable with currently existing observational means. For some
planets, we might have a previous knowledge of the geostationary
altitude (see section~\ref{sec:other}). Since nature has no particular
preference for this orbit, the mere detection of a population of
objects at precisely the CEB altitude would be extremely suggestive of
an artificial origin.

\section{The Clarke exobelt}
\label{sec:ceb}

Artificial satellites in geostationary and geosynchronous orbits are
useful to us for a number of purposes, including telecommunications,
surveillance, wildfire control, geolocation, spionage, wildlife
tracking as well as other scientific studies and civil or military
applications. The geostationary orbit, often named after Clarke, who
explored its practical usefulness for communication purposes
(\citealt{C45}), is specially interesting because satellites placed
there will remain static as seen from the ground reference
frame. However, the available space in that orbit is limited. A
moderately advanced civilization might eventually populate it with a
relatively high density of objects, making it advisable (cheaper in a
supply-and-demand sense) to use geosynchronous orbits when possible
for those satellites whose requirements are less strict and allow for
some degree of movement along the North-South direction on the
sky. Over time, one might expect that societal needs would eventually
drive an increase of object density in a band around the geostationary
orbit, forming a CEB.
%All these
%objects form the CEB, named after \citet{C45}, who explored the
%potential applications of such orbits for telecommunications.

%% The "ht!" tells LaTeX to put the figure "here" first, at the "top" next
%% and to override the normal way of calculating a float position
\begin{figure}
\includegraphics[width=0.5\textwidth]{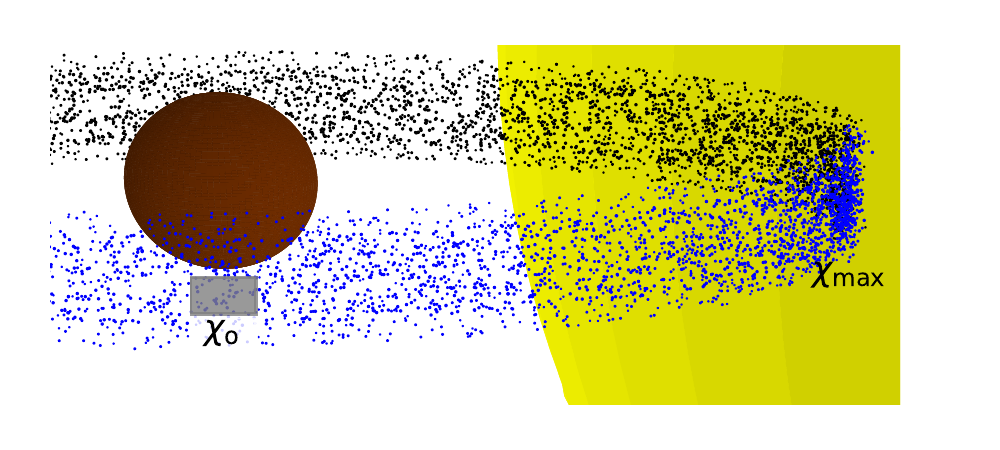}
\caption{
  \label{fig:sketch}
  Illustrative sketch of a planet with a CEB transiting its
  parent star. The size and surface density of the individual objects
  has been exaggerated for better visualization. The face-on opacity
  \chio \, (see text for definition) is 0.2 in this case. The edge
  opacity \chimax \, reaches its maximum possible value of 1.}
\end{figure}

Figure~\ref{fig:sketch} shows a sketch of a transiting exoplanet with
a CEB for illustration. The size of individual objects and the belt
opacity have been exaggerated\footnote{in comparison with the
  simulations presented in this paper} in the figure for
visibility. The face-on opacity \chio \, (the opacity of a small
surface element when viewed from a perpendicular direction) is 0.2 in
this example. The line-of-sight (hereafter l.o.s.) opacity increases
as we move away from the center, reaching a maximum value \chimax \,
at the edge.

\begin{figure}
  \vspace{.01in}
  \hspace{-.11in}
\includegraphics[width=0.5\textwidth]{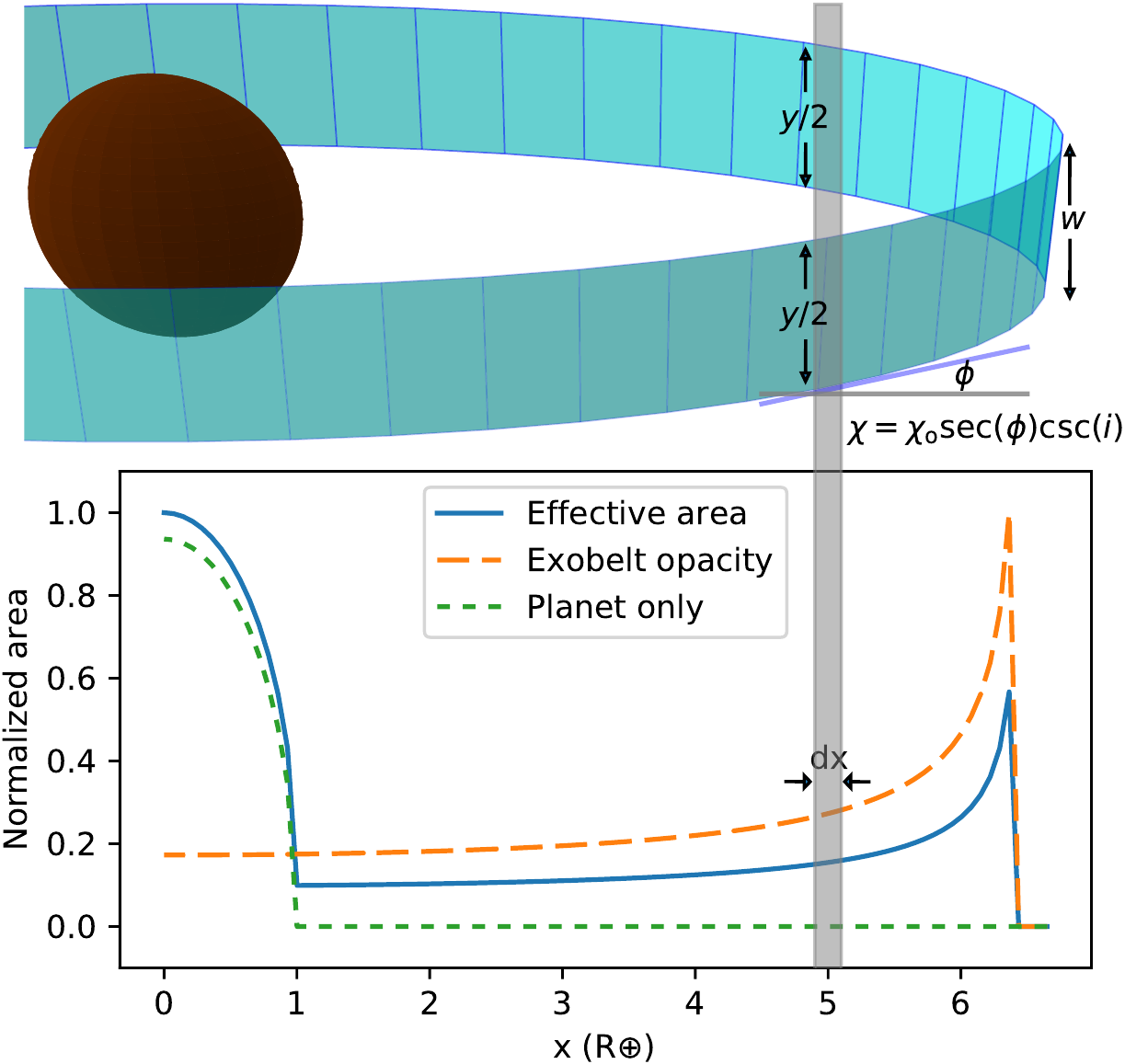}
\caption{
  \label{fig:model}
  Upper panel: Numerical model of a CEB. The azimuth discretization is
  deliberately coarse for better visualization. The actual
  calculations have a much finer discretization. See text for symbol
  definitions. Lower panel: Solid blue line: Effective area
  $\alpha(x)$ presented by the entire system as a function of
  $x$. Dashed orange line: Opacity $\chi$ of the belt in the observer
  direction, normalized to its maximum value. \chio \, has been set to
  0.03 for an adequate visualization of the belt contribution in this
  figure (it is much lower in the simulations shown in the figures
  below). Dashed green line: Same as the solid blue line if the planet
  had no CEB. Both panels share the same abscissae to allow for a
  direct comparison.}
\end{figure}

The CEB models presented here are characterized by the following
parameters (see Fig~\ref{fig:model}, upper panel): radius ($r_C$),
width ($w$), face-on opacity (\chio) and inclination of the equatorial
plane with respect to the plane of the sky ($i$). A CEB seen edge-on
would have $i=90$\degree . For simplicity, the belt has well-defined
sharp borders. It is modeled as a continuous surface, discretized in
narrow rectangles (surface elements), as seen in Fig~\ref{fig:model}.
Also for simplicity, the surface is perfectly cylindrical and has zero
geometrical thickness. This implies that the model does not consider
objects with eccentric orbits and all objects are assumed to have
exactly the same orbital altitude. These approximations are in good
agreement with Earth's current satellite population. According to
publicly available data (see Section~\ref{sec:earthbelt} below), the
vast majority of geosynchronous satellites have nearly circular orbits
(less than 2\% have eccentricities above 0.01). The spread in altitude
is of $\simeq$150~m, only a few parts per million (ppm) of the belt
radius.

The belt opacity (\chio ) is defined as the fraction of light in our
l.o.s. that is blocked by a surface element (the vertical rectangles in
Fig~\ref{fig:model}). It is a parameterization of the surface density
and size of objects on the belt. If a surface element is tilted with
respect to our l.o.s., we would see a higher opacity:
\begin{equation}
  \label{eq:chi}
  \chi=\chi_{\mathrm{o}} \sec(\phi) \csc(i) \, .
\end{equation}
The width parameter $w$ is related to the maximum orbital inclination
$\gamma$ of geosynchronous satellites $w=r_C \sin(\gamma)$. The
current industry standard\footnote{See, e.g., the Inter-Agency Space
  Debris Coordination Committee guidelines at
  \url{http://www.unoosa.org/documents/pdf/spacelaw/sd/IADC-2002-01-IADC-Space_Debris-Guidelines-Revision1.pdf}}
defines a protected geosynchronous region delimited by
$\gamma=15$\degree . According to publicly available databases (see
Section~\ref{sec:earthbelt} below), 97\% of currently active
geosynchronous satellites have orbital inclinations in this
range. Taking this value as a starting point, we shall explore
simulations with $\gamma$ between 15 and 30\degree.

In order to determine the imprint of a CEB on the light curve of the
star, we need to calculate the amount of light blocked by the system
at each point as it moves across the stellar disk. Taking the $x$-axis
as the direction of the planet motion on the sky, Fig~\ref{fig:model}
(lower panel) plots the effective area $\alpha(x)$, defined as the
geometrical area of the system multiplied by its opacity:
\begin{equation}
  \label{eq:alpha}
  d\alpha(x) = y(x) \chi(x) dx \, ,
\end{equation}
(where $y$ is the projected size of the system in the direction
perpendicular to $x$ on the sky). The upper and lower panels in
Fig~\ref{fig:model} share the same abscissae and may be directly
compared. The belt opacity varies as indicated by the dashed curve. It
has a minimum value \chio \, at $x=0$, where the belt orientation is
most perpendicular to the l.o.s., and increases as the inverse of the
cosine of the surface element angle $\phi$. The surface elements are
represented in the upper panel. They are shown very coarsely for
visualization but the actual numerical model used for the calculations
has a much finer discretization. For each interval $dx$, the code
computes the projected area, its opacity and the amount of overlap
between the front side and the far side of the belt. If there is
overlap (as for values of $x > 5.5$ in the figure), their opacities
are added. Since we are working in the \chio $\ll 1$ regime, all
the results in this paper will scale with \chio .

\section{Simulations}
\label{sec:simulations}

\subsection{Earth's belt}
\label{sec:earthbelt}

Let us first consider the Earth-Sun system as seen from another
star. At the present time, we have too few satellites and debris to be
detectable at interstellar distances with the technique proposed here.
It is difficult to determine with precision the amount of objects
in our Clarke belt. Publicly available databases are incomplete and do
not consider classified satellites, dead or decommissioned devices,
space junk, etc. Nevertheless, it is still insightful to calculate
a rough order-of-magnitude estimate of our current \chio .

A particularly useful database is the compilation of data from public
sources made by the Union of Concerned Scientists (UCS)\footnote{The
  UCS database with references to the original sources is available
  online at \url{http://www.ucsusa.org/satellite_database}. The
  calculations presented in this paper make use of release
  9-1-17.}. Currently, the list contains parameters for 1738
satellites, of which approximately one third are in geostationary or
geosynchronous orbits. Assuming a typical radius of 1~m, we have that
\chio$\simeq 3 \times 10^{-13}$.

\begin{figure}
\includegraphics[width=0.5\textwidth]{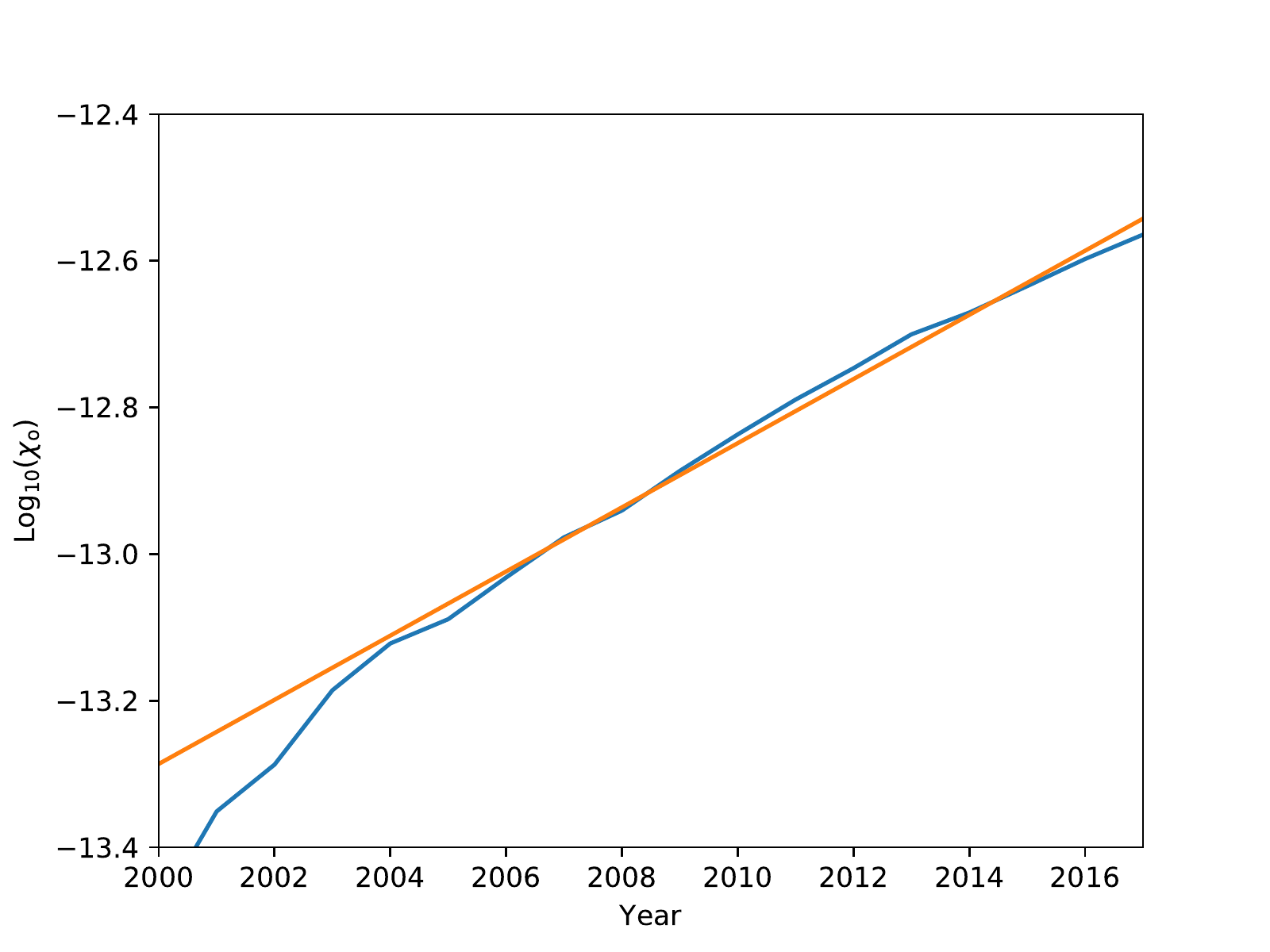}
\caption{ Blue: Variation of the \chio \, parameter for the Clarke
  belt of humanity over the last decades (notice that the ordinate
  axis is logarithmic), assuming a typical object radius of 1~m and
  using public satellite data. Orange: Fit to an exponential increase.
  \label{fig:earth}
}
\end{figure}

In order to become visible from nearby stars with our current
observational capabilities, the Clarke belt of our planet would need
to be about \chio$\sim$10$^{-4}$ (see below). Therefore, it is
reasonable to assume that we are probably orders of magnitude below
the detection threshold of any possible observer. However, our belt is
becoming increasingly populated. Figure~\ref{fig:earth} shows that the
Earth belt opacity \chio \, has been growing exponentially over the
last 15 years. If this trend is extrapolated into the future, we would
reach the ``observable'' threshold ($\sim 10^{-4}$) around the year
2200.

Obviously, this extrapolation should not be viewed as a
prediction. There is no reason to assume that the current exponential
growth will be sustained for another 200 years. It might slow down if
the demand for orbital devices were to decline, or it could accelerate
if new technologies were developed that either require or facilitate
the addition of more devices. In this respect it is worth pointing out
another Clarke invention: a ``space elevator'' system would
tremendously facilitate access to geostationary orbit, which is a
natural place to stop, and would likely speed up the rate of \chio \,
growth. In summary, the 2200 date is not even a rough guess of when
humanity will reach detectability threshold but rather an indication
that this outcome is a reasonable expectation for the near future,
given current trends.

\begin{figure}
\includegraphics[width=0.5\textwidth]{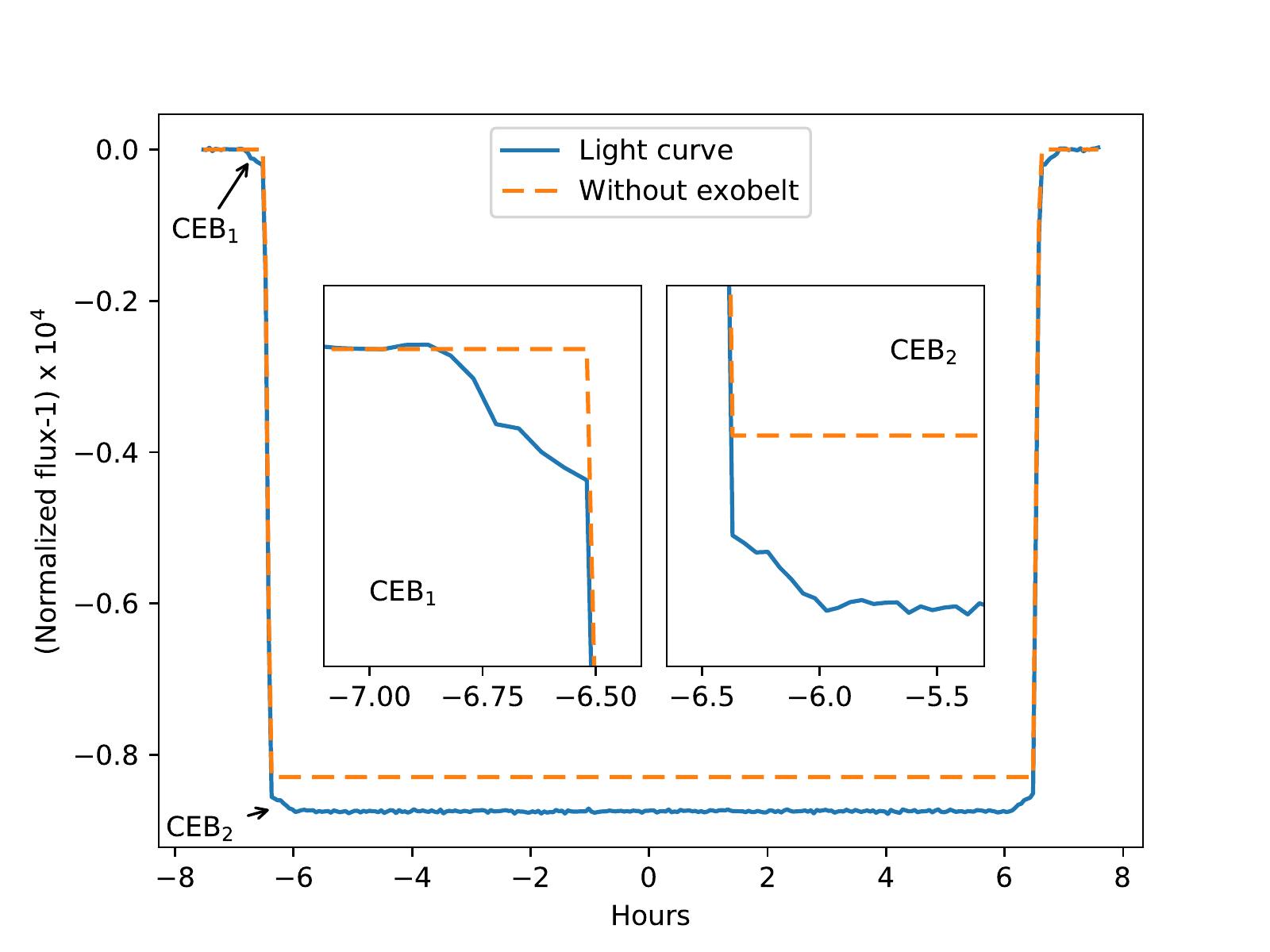}
\caption{Synthetic light curve computed for an Earth-like planet
  transiting a Sun-like star at a distance of 10 light-years. The
  planet has a CEB of \chio = 10$^{-3}$, $\gamma = 20$\degree \, and
  $i = 70$\degree. The arrows indicate the positions of the belt transits
  during the ingress (they have symmetric counterparts during
  egress). CEB$_1$ marks the light-curve drop from the CEB exterior
  ingress until the planet exterior ingress. From CEB$_1$ to CEB$_2$
  the planet moves into the stellar disk. CEB$_2$ marks the ingress of
  the trailing half of the CEB behind the planet. The inset plots show
  close ups of both CEB$_1$ and CEB$_2$ events.
  \label{fig:lcearth}
}
\end{figure}

Let us consider the appearance of our system, as seen from another
star, in a hypothetical future with a more cluttered Clarke
belt. Figure~\ref{fig:lcearth} depicts the light curve of Earth with a
\chio$=10^{-3}$ belt, transiting the Sun as viewed from a distance of
10 light-years. Random noise is computed assuming a perfect 10~m
aperture telescope doing white-light photometry with 60~s
exposures. This simulation assumes a black-body spectral distribution
of the stellar radiation, filtered with the spectral response of the
Kepler
mission\footnote{\url{https://keplergo.arc.nasa.gov/CalibrationResponse.shtml}}. All
of the light curves computed in this paper have noise dominated by
photon detection statistics.  For simplicity, the code assumes zero
impact parameter in all transits and neglects limb-darkening. The
events marked as CEB$_1$ and CEB$_2$ in the plot (see figure insets
for details) are produced by the belt. Both events are clearly above
the noise, at the ppm level. For a sake of reference, consider that
the Kepler mission has a photometric precision of $\sim$10~ppm.
Therefore, in this case study, the belt would be detectable using
current technology. It is not far from the capabilities of existing
planet hunting missions.

\subsection{Other planets}
\label{sec:other}

The radius $r_C$ of a CEB is a function of the planet mass but also
its rotation, according to the expression:
\begin{equation}
  \label{eq:r}
r^3 = {G \over {4 \pi^2}} M T^2 \,
\end{equation}
where $G$ is the universal gravitational constant, $M$ is the mass of
the planet and $T$ is the rotation period. An interesting insight from
the above formula is that $r_C$ depends very weakly on $M$, only as a
power of $1/3$. This is very fortunate because, for many transiting
exoplanets, we have no radial velocity measurements and therefore the
mass is uncertain. However, we do have an accurate determination of
the size and our uncertainty in the planet density translates into a
much lower uncertainty in the CEB radius $r_C$. For instance, Venus,
Earth and Mars have a density spread of 37\% (Mercury is peculiar,
with an anomalously high density more characteristic of a planetary
core). If we take this as the typical uncertainty on Earth-like planet
densities, the associated uncertainty on $r_C$ would be 11\%. This
discussion does not apply to super-Earths, which are likely to have
considerably higher densities.

The rotation period $T$ may be straightforward, very difficult or
impossible to estimate, depending on the situation. \citet{BK17}
examined the observational requirements to produce time-dependent
spectroscopic maps of a planet surface. For the case of Proxima b they
concluded that a low scattered light telescope with an aperture of
more than 12~m may be able to accomplish this task. Surface
inhomogeneities would imprint a periodic spectral modulation that
would reveal its rotation period with very high accuracy. If a planet
exhibits surface albedo variations on a global scale, it would be
possible to determine its rotation period more easily, and therefore
at greater distances, than via spectroscopy. \citet{KJ10} showed that
future space missions may produce surface maps of planets up to 5~pc
away. In any case, both photometric and spectroscopic rotation
measurements would require next-generation instrumentation and would only
be feasible for systems relatively close to us.

The planet rotation period might be straightforwardly obtained if it
is tidally locked, which is probably very frequent for the most
interesting exoplanet candidates. Earth seems to be rather peculiar in
this regard. According to some models (\citealt{B17}), without
atmosphere, moon and assuming constant tidal properties, it would have
already synchronized its rotation with the orbital motion around the
Sun, resulting in days of the same duration as a year. However, tidal
locking in the habitable zone (\citealt{KWR93}; \citealt{K13}) may be
very common (however, counterarguments exist for planets with dense
atmospheres, e.g. \citealt{ADLM+17}) and its frequency should increase
from G to K and M type stars. The smaller stars (types K and M) are by
far the most abundant in the galaxy and, furthermore, exoplanet
detection is observationally easier there. Therefore, they have become
the prime candidates for planet search projects and, lacking better
criteria, possibly the most interesting targets for the search of
indicators such as the one explored in this paper. According to
\citet{B17}, half of the Kepler planet candidates and the vast
majority of the ones expected to be discovered by TESS become tidally
locked in less than 1~Gyr. Therefore, it seems plausible that we might
be able to obtain a good estimate of the CEB radius for a potentially
interesting candidate.

\begin{figure}
\includegraphics[width=0.5\textwidth]{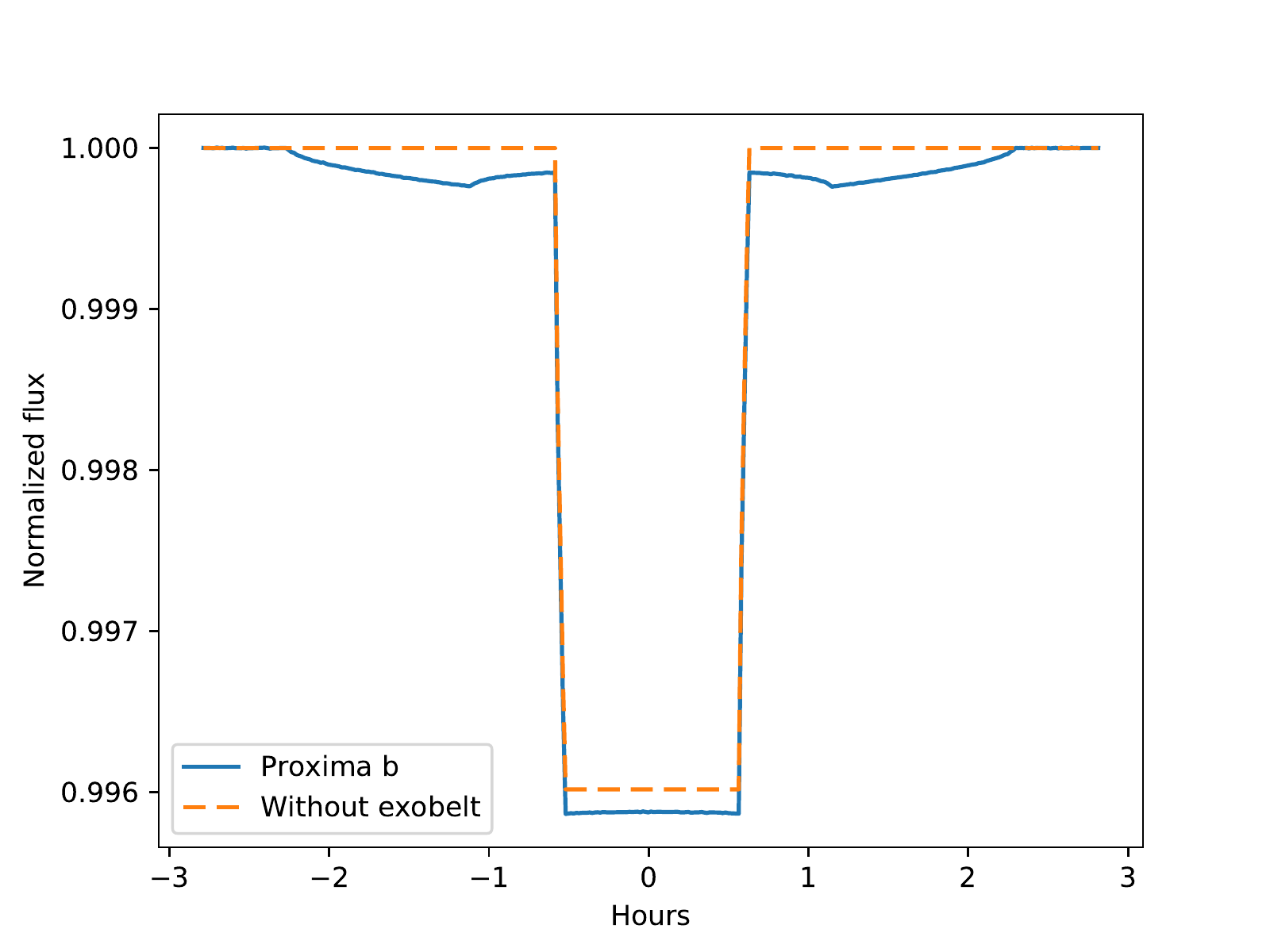}
\caption{Synthetic light curve computed for Proxima~b with a CEB
  having \chio = 5$\times$10$^{-5}$, $\gamma = 20$\degree,
  $i=80$\degree.
  \label{fig:lcproximab}
}
\end{figure}

Let us now consider the light curve, starting with the closest
exoplanet in habitable zone, Proxima~b, as a particularly relevant
reference. Figure~\ref{fig:lcproximab} shows the light curve of a
transit with \chio = 5$\times$10$^{-5}$, $\gamma = 20$\degree \, and
$i = 80$\degree \, observed with a 10~m telescope in near-infrared
J-band photometry (using the standard Mauna Kea filter
definition). The star and planet data employed are listed in
Table~\ref{table:planets} and have been obtained from \citet{AEAB+16}
and \citet{BA17}.  As the figure shows, we could easily detect the
presence of a CEB with \chio \, lower than 10$^{-4}$.

\begin{figure}
\includegraphics[width=0.5\textwidth]{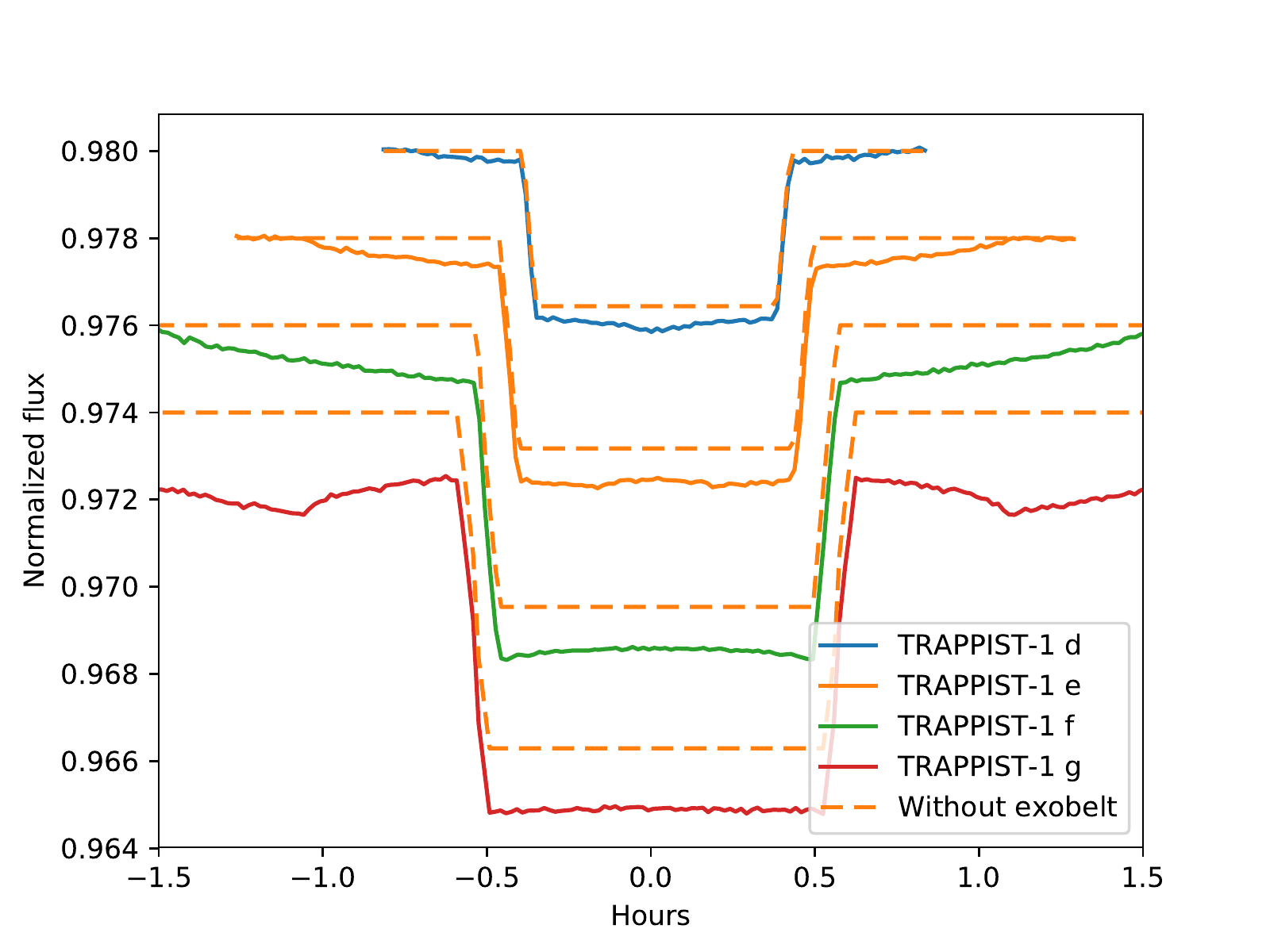}
\caption{Synthetic light curve computed for TRAPPIST-1 planets with a
  CEB having \chio = 5$\times$10$^{-4}$, $\gamma = 20$\degree,
  $i=80$\degree. The light curves for the various planets have been
  shifted vertically to show them all without overlap.
  \label{fig:lctrappist1}
}
\end{figure}

Another system of great interest is TRAPPIST-1, with seven confirmed
rocky planets in habitable zone or very close to it, allowing
for an accurate determination of their masses and almost certainly
tidally locked. The most likely habitable planets in the TRAPPIST-1
system are d to g, which span a range of masses, distances and periods
listed in Table~\ref{table:planets}. Data for the star and planets
have been taken from \citet{vGFG+17}, \citet{WWB+17} and
\citet{DGT+18}. The synthetic light curves of all of these planets
with a CEB are shown in Fig~\ref{fig:lctrappist1}. The parameters
chosen for the calculations are \chio = 5$\times$10$^{-4}$, $\gamma =
15$\degree \, and $i = 80$\degree . Reducing \chio \, to 10$^{-4}$
would still result in observable signatures for all except the
innermost planet TRAPPIST-1~d.

\begin{deluxetable*}{l|cccccccc}
  \tablecaption{Planet parameters employed for light-curve calculations. In parentheses are the actual values published in the references, when different or highly uncertain.
  \label{table:planets}}
  \tablehead{
    \colhead{Planet} & \colhead{Mass} & \colhead{Radius} & \colhead{Orbit} & \colhead{Rotation} & \colhead{Distance} & \colhead{Star radius} & \colhead{Star luminosity} & \colhead{T$_\mathrm{eff}$} \\
    \colhead{}      & \colhead{(M$_\oplus$)} & \colhead{(R$_\oplus$)}  &
    \colhead{(UA)}   &   \colhead{(days)}       &    \colhead{(ly)}    &
    \colhead{(R$_\odot$)}   &  \colhead{(10$^{-4}$ $\times$ L$_\odot$)} & \colhead{(K)}
  }
%  \colnumbers
  \startdata
  Proxima b      & 2.6 ($>$1.3)  & 1.0 (0.8-1.5)   & 0.0485 & 11.2 & 4.22 & 0.14 & 15 & 3050 \\
  TRAPPIST-1 d   & 0.33        & 0.78   & 0.022 & 4.0   & 39.6 & 0.12 & 5.2 & 2516 \\
  TRAPPIST-1 e   & 0.60        & 0.91   & 0.029 & 6.1   & 39.6 & 0.12 & 5.2 & 2516 \\
  TRAPPIST-1 f   & 0.70        & 1.05   & 0.038 & 9.2   & 39.6 & 0.12 & 5.2 & 2516 \\
  TRAPPIST-1 g   & 1.30        & 1.15   & 0.047 & 12.3  & 39.6 & 0.12 & 5.2 & 2516 \\
  \enddata
\end{deluxetable*}

\subsection{Rings}
\label{sec:rings}

\citet{WCZ+17} explored the possible confounding natural factors in
the general context of artificial constructions around
stars or planets, including rings. At first sight, the imprint of a
CEB on the light curve is similar to that of a ring system. A smaller
dip in intensity, marked as CEB$_1$ in Fig~\ref{fig:lcearth}, appears
just before the planet exterior ingress (the intensity drop caused by
the planet moving into the stellar disk). Afterwards, another dip
(marked as CEB$_2$ in Fig~\ref{fig:lcearth}) appears just after the
interior ingress (the endpoint of the intensity drop). This pattern of
dips just before and after the planet ingress is also produced by ring
systems (\citealt{AS04}). The question then is, can we infer from this
signature the presence of extraterrestrial intelligence? Let us assume
that we detect a promising candidate planet, meeting some basic
criteria (rocky and at a suitable temperate distance from
its parent star) with dips before and after planet ingress
corresponding to a distance $x$ from its center.

The first question is whether planets of this kind are likely to host
ring systems. In principle, dynamical studies suggest that rings might
indeed exist around temperate planets (\citealt{SC17}) and remain
stable on Gyr time-scales. However, there are also indications that
rings might occur predominantly in planets beyond the ice line and
would be less frequent in the habitable zone rocky planets that are
the main target of current SETI. The search for exoplanets, which is
biased towards inner planets with shorter orbital periods, has not yet
found any proper ring system in the more than 3,700 planets discovered
thus far. The only exception, 1SWASP J140747.93–394542.6
(\citealt{MQP+12}), has a disk that is much larger than the planet's
Roche lobe and should probably be considered a transient
protosatellite moon rather than a stable ring system
(\citealt{HBA+18}). Furthermore, in our own Solar System we find that
all the planets beyond the ice line have rings, in addition to other
smaller bodies such as Haumea, Chariklo and, possibly, Chiron. In
contrast, none of the inner Solar System bodies have
rings. \citet{H15} proposed that this could be explained assuming that
the ice-rich material becomes weak at low temperatures ($\sim$70~K),
facilitating fragmentation and formation of rings.

We still have very little understanding of exoring formation and their
probability of occurrence, especially in the habitable zone. Given the
considerations discussed above, the observation of CEB-like features
in the transit of a candidate planet should be viewed as extremely
suggestive. The next obvious question would be whether it has the right
orbital altitude.

For most planets, we can determine the CEB radius $r_C$ directly from
their mass and rotation period. Recall that, as discussed above, $r_C$
is very weakly dependent on the mass (it goes with the power of
1/3). This means that, even in the absence of proper mass
measurements, the size of an Earth-like planet (which is
straightforwardly determined from the transit light curve) is
sufficient to constrain $r_C$ to approximately 11\%. Rotation might be
measured observationally (although this would be very challenging) or
derived trivially if the planet is tidally locked, which may be the
case for many interesting candidates. With these considerations in
mind, it is very likely that we would have a robust determination of
$r_C$ for our planet candidate. If it happened to exhibit the dips at
the right distance ($x=r_C$), it would be a very strong indication of
an artificial origin. Geostationary orbits are very interesting for a
society but are not preferred by any known natural process.

Let us now examine the issue of whether detailed extensive
observations could resolve the ambiguity between rings and a
CEB. These two structures have a different intrinsic geometry. Both
are extremely thin and flat but in different (perpendicular)
directions. Rings are extended in the radial direction and thin in
inclination. CEBs, on the other hand, are thin in the radial direction
and extended in inclination. With some straightforward modifications,
the code used for the calculations presented in the previous sections
may be adapted to compute a simple ring system in the same conditions,
allowing for a detailed comparison of both scenarios. Note, however,
that a large ring system might have a more complex geometry with
radial variations of opacity, such as the gaps in Saturn's rings,
which would not be captured by this simple model.

\begin{figure}
  \vspace{0.1in}
\includegraphics[width=0.5\textwidth]{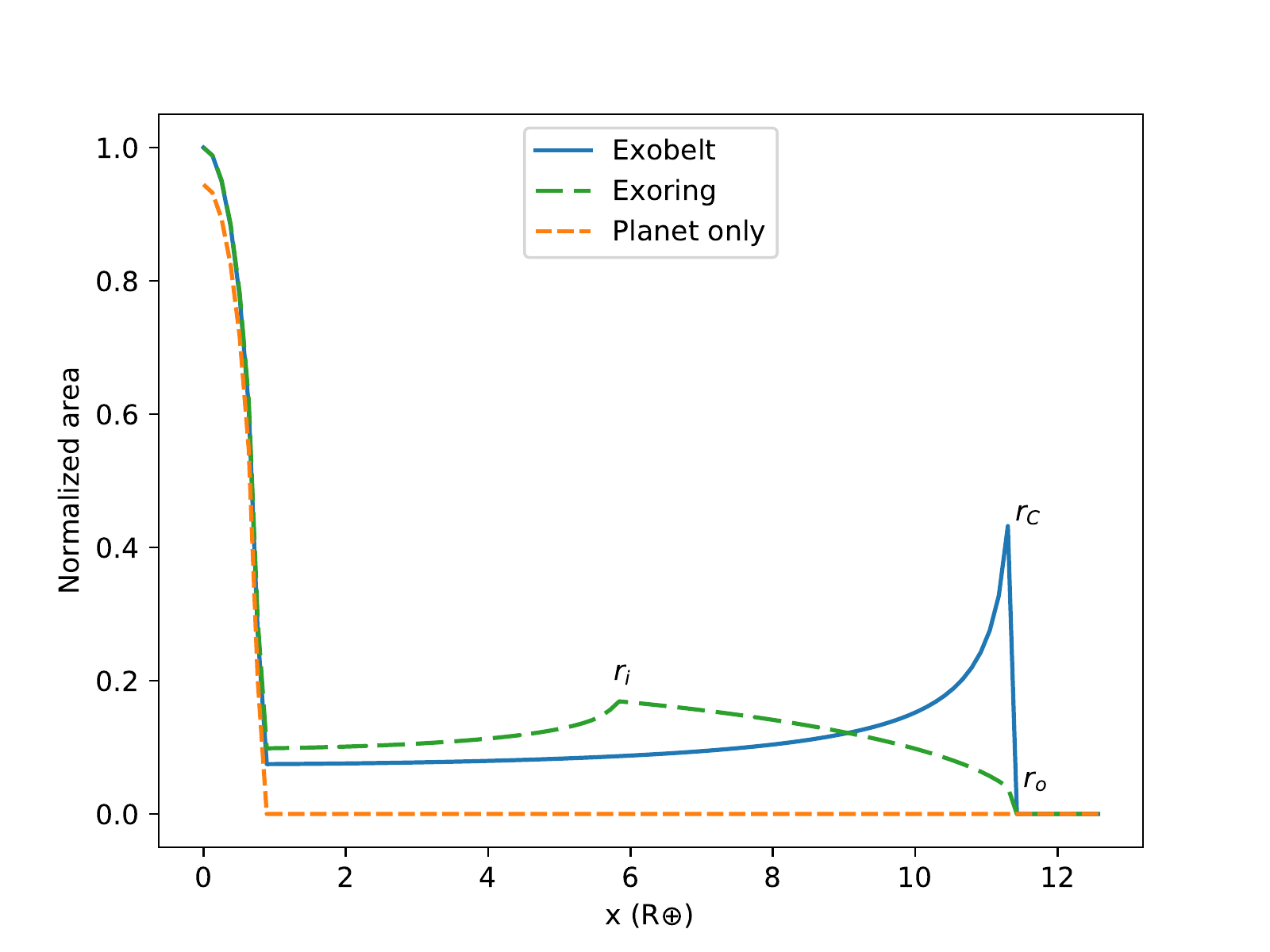}
\caption{Effective area $\alpha(x)$ presented by a planet with CEB
  (solid blue), a planet with rings (dashed green) and an isolated
  planet (dashed orange). Notice the overall concave shape of the CEB
  curve as opposed to the ring, which changes from concave to convex
  from left to right at $r_i$. This simulation has $i=80$\degree \,
  and $\gamma=20$\degree . The ring has inner and outer radii
  $r_i=r_C/2$ and $r_o=r_C$.
  \label{fig:ring1}
}
\end{figure}

As a result of their different geometry, the effective area curve
$\alpha(x)$ produced by rings and CEBs are markedly different. A
comparison is plotted in Fig~\ref{fig:ring1}. The figure illustrates a
particular configuration but the overall conclusion is rather general
and easily understood from the geometry of the problem. The CEB has its
maximum opacity at the very edge but its projected area does not
change significantly with $x$. The ring, on the other hand, has an
approximately constant opacity and its area increases gradually from
the outer to the inner radius.

\begin{figure}
  \includegraphics[width=0.5\textwidth]{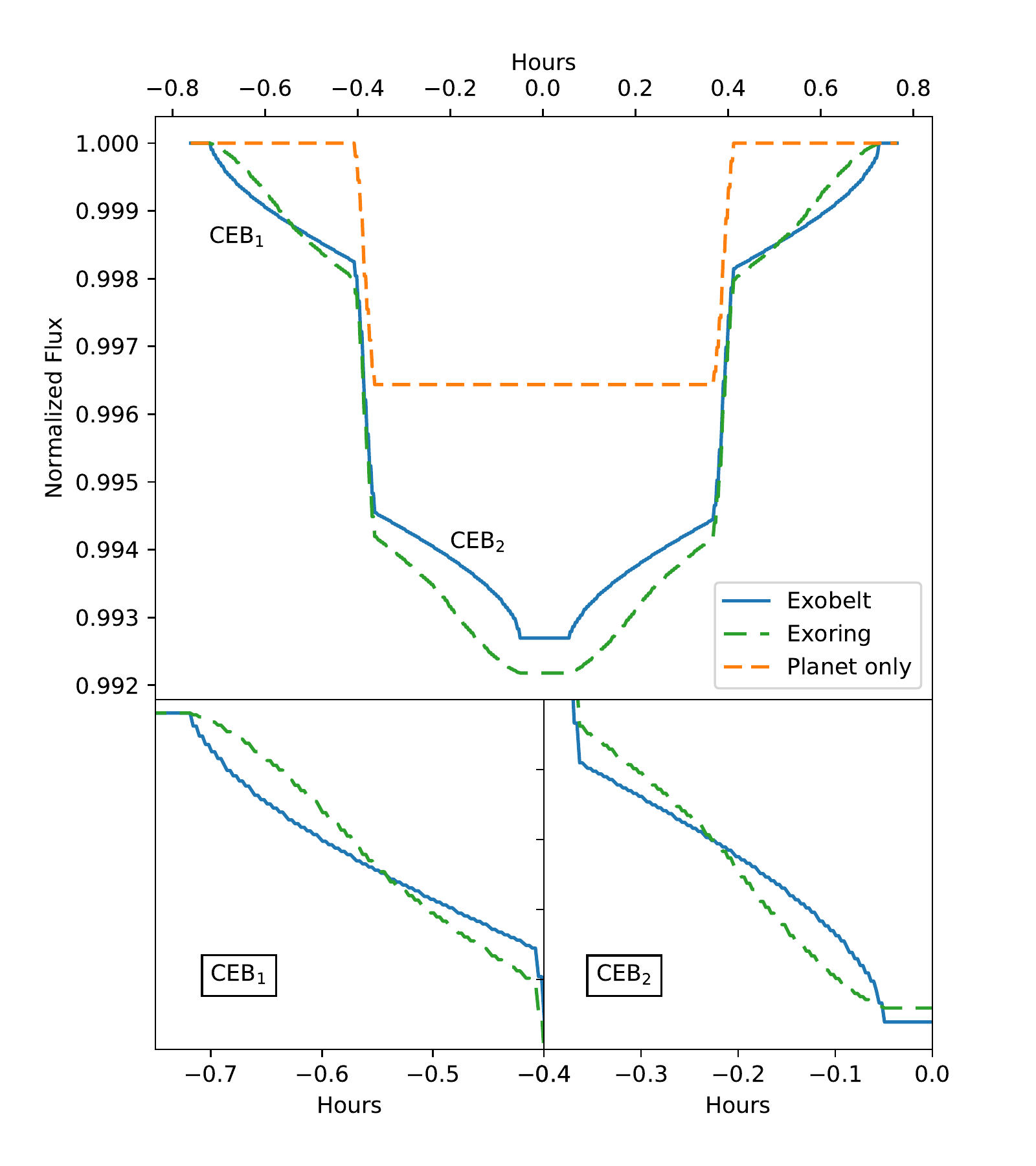}
  \caption{Synthetic light curves corresponding to the simulation of
    Fig~\ref{fig:ring1}. The upper panel shows the entire transit. The
    lower panels show a close-up of the regions CEB$_1$ and
    CEB$_2$. The green dashed line in CEB$_2$ has been shifted
    vertically to facilitate visual comparison.
  \label{fig:ring2}
}
\end{figure}

The differences in $\alpha(x)$ translate into subtle but measurable
differences in the light curve. This is shown in
Fig~\ref{fig:ring2}. For a CEB, the first dip (CEB$_1$) is always
concave, while the second one (CEB$_2$) is always convex. This is a
direct result of the variation of $\chi$ in the transit direction
which, as explained above, goes with the secant of $\phi$. The ring, on
the other hand, switches from convex to concave in both dips. This
behavior is a consequence of the variation in the ring projected area,
from edge to center (see Fig~\ref{fig:ring1}). For both, CEB and ring
system, the light curve convexity reflects the convexity of
$\alpha(x)$, which is constant for the CEB (always concave) but
switches sign for the ring system, going from convex at the outer
radius to concave at the inner radius.

Another difference is that the ring system has a flat start and end to
the dip. The transition between a flat light curve and the dip is
smooth. The CEB, on the other hand, has a maximum slope at the start
and the end. There is a marked discontinuity in the derivative of the
light curve at the endpoints. This is caused by the fact that the CEB
has its maximum opacity at the edge and then drops abruptly to
zero. The ring, on the other hand, starts with zero area at the outer
radius and then increases gradually as the ingress progresses
(Fig~\ref{fig:ring1}).

\section{Conclusions}

One of the main conceptual difficulties in the SETI effort is our
complete lack of knowledge on the motivations or interests of other
civilizations. We are forced to guess what other peoples, with whom we
share absolutely no common cultural background, are trying to build or
accomplish. For instance, our searches are focused on planets but
perhaps civilizations with highly advanced capabilities might have
decided to move away from the perils of stars and planets and dwell
instead in the dark safety of interstellar space, connected to their
parent system only to retrieve minerals and fusion fuel. This is
simply an example of the many uncertainties and unconscious biases
involved in the complex SETI guesswork. Therefore, it is important to
develop technomarkers with which to expand the search to include
civilizations at a level of technological development as close as
possible to our own.

The indicator analyzed here is a step in that direction. A CEB does
not require of any technology that we do not have, only a more
extensive use of orbital space. Perhaps their civilization is older
than ours and has had more time to populate it. Or perhaps it has been
driven by a stronger push for space devices, for reasons that we could
only speculate about.

The numerical model has some basic simplifying assumptions. In
reality, there would be more devices in other orbits around the
planet. The belt might have some thickness due to slightly different
orbital altitudes and some fraction of geosynchronous satellites may
have eccentric orbits. However, small deviations from this idealized
model will not have a significant effect on the main results. There is
always the possibility that an ETI might develop a completely
different approach in populating their CEB but the goal of this paper
is to explore the consequences of a direct extrapolation of our
current trends.

The simulations presented here show that CEBs may in some situations
be detectable with existing instrumentation. The best candidates are
planets around red dwarfs in tidal locking, in line with the optimal
conditions for habitable exoplanet search. An initial difficulty would
be how to distinguish between a CEB and a ring system. However, once a
candidate has been identified, detailed follow-up observations may
resolve this ambiguity from the shape of the light curve. In any case,
the detection of a dense belt of objects at the distance of
geostationary orbit would be a very strong evidence for the presence
of ETI, especially considering that rings around habitable rocky
planets are probably rather uncommon.

While the similarity between a CEB and a ring system poses an initial
difficulty, it also opens new opportunities. Existing interest in the
physics of exorings and exomoons means that large efforts will be
devoted in future photometric missions to examine rocky planet
transits for evidence of such objects. This paper shows how future
positive detections of orbital material may be further scrutinized for
evidence of CEBs, making the search for moderately advanced
technologies ``piggyback'' on such missions.

The total mass of the entire belt for all the cases considered here
(Earth, Proxima b and TRAPPIST-1 planets d to g) is between~10$^{12}$
and~10$^{14}$~kg, assuming \chio=5$\times$10$^{-4}$,
$\gamma$=20\degree \, with average object radius and mass of 1~m and
100~kg. This range is between the mass of a comet and that of a
mountain. It is not an unreasonable requirement for a moderately
advanced civilization.
%The amount of mass in currently operational
%motor vehicles on Earth is of that order.

One exciting perspective about a CEB discovery is that it would most
likely point to the presence of an active civilization. Other
technomarkers, such as Dyson spheres or swarms, could have been built
by species that disappeared, moved away or became extinct long ago. A
crowded CEB, on the other hand, requires active maintenance to keep
objects in proper orbits and away from collisions with other nearby
objects. A dynamical study of the relevant timescales is beyond the
scope of this paper but for our current satellites, such scales are
typically of the order of decades.

As with any other technomarkers, the search for CEBs is a long shot. We
have no idea if they exist or how likely they are to occur. However,
given that candidate identification is based on the same light curve
observations that are currently demanded in the search for habitable
exoplanets, it does not require of any additional effort, at least
initially, other than being alert for possible detections.




%\begin{thebibliography}{}

  \bibliography{paper}

\begin{thebibliography}{}
\expandafter\ifx\csname natexlab\endcsname\relax\def\natexlab#1{#1}\fi
\providecommand{\url}[1]{\href{#1}{#1}}

\bibitem[{{Angel} {et~al.}(2006){Angel}, {Codona}, {Hinz}, \& {Close}}]{ACH+6}
{Angel}, R., {Codona}, J.~L., {Hinz}, P., \& {Close}, L. 2006, in \procspie,
  Vol. 6267, Society of Photo-Optical Instrumentation Engineers (SPIE)
  Conference Series, 62672A

\bibitem[{{Anglada-Escud{\'e}} {et~al.}(2016){Anglada-Escud{\'e}}, {Amado},
  {Barnes}, {Berdi{\~n}as}, {Butler}, {Coleman}, {de La Cueva}, {Dreizler},
  {Endl}, {Giesers}, {Jeffers}, {Jenkins}, {Jones}, {Kiraga}, {K{\"u}rster},
  {L{\'o}pez-Gonz{\'a}lez}, {Marvin}, {Morales}, {Morin}, {Nelson}, {Ortiz},
  {Ofir}, {Paardekooper}, {Reiners}, {Rodr{\'{\i}}guez},
  {Rodr{\'{\i}}guez-L{\'o}pez}, {Sarmiento}, {Strachan}, {Tsapras}, {Tuomi}, \&
  {Zechmeister}}]{AEAB+16}
{Anglada-Escud{\'e}}, G., {Amado}, P.~J., {Barnes}, J., {et~al.} 2016, \nat,
  536, 437

\bibitem[{{Arnold} \& {Schneider}(2004)}]{AS04}
{Arnold}, L., \& {Schneider}, J. 2004, \aap, 420, 1153

\bibitem[{{Astropy Collaboration} {et~al.}(2013){Astropy Collaboration},
  {Robitaille}, {Tollerud}, {Greenfield}, {Droettboom}, {Bray}, {Aldcroft},
  {Davis}, {Ginsburg}, {Price-Whelan}, {Kerzendorf}, {Conley}, {Crighton},
  {Barbary}, {Muna}, {Ferguson}, {Grollier}, {Parikh}, {Nair}, {Unther},
  {Deil}, {Woillez}, {Conseil}, {Kramer}, {Turner}, {Singer}, {Fox}, {Weaver},
  {Zabalza}, {Edwards}, {Azalee Bostroem}, {Burke}, {Casey}, {Crawford},
  {Dencheva}, {Ely}, {Jenness}, {Labrie}, {Lim}, {Pierfederici}, {Pontzen},
  {Ptak}, {Refsdal}, {Servillat}, \& {Streicher}}]{ACRT+13}
{Astropy Collaboration}, {Robitaille}, T.~P., {Tollerud}, E.~J., {et~al.} 2013,
  \aap, 558, A33

\bibitem[{{Auclair-Desrotour} {et~al.}(2017){Auclair-Desrotour}, {Laskar},
  {Mathis}, \& {Correia}}]{ADLM+17}
{Auclair-Desrotour}, P., {Laskar}, J., {Mathis}, S., \& {Correia}, A.~C.~M.
  2017, \aap, 603, A108

\bibitem[{{Barnes}(2017)}]{B17}
{Barnes}, R. 2017, Celestial Mechanics and Dynamical Astronomy, 129, 509

\bibitem[{{Berdyugina} \& {Kuhn}(2017)}]{BK17}
{Berdyugina}, S.~V., \& {Kuhn}, J.~R. 2017, ArXiv e-prints, arXiv:1711.00185

\bibitem[{{Bixel} \& {Apai}(2017)}]{BA17}
{Bixel}, A., \& {Apai}, D. 2017, \apjl, 836, L31

\bibitem[{{Chennamangalam} {et~al.}(2017){Chennamangalam}, {MacMahon}, {Cobb},
  {Karastergiou}, {Siemion}, {Rajwade}, {Armour}, {Gajjar}, {Lorimer},
  {McLaughlin}, {Werthimer}, \& {Williams}}]{CMC+17}
{Chennamangalam}, J., {MacMahon}, D., {Cobb}, J., {et~al.} 2017, \apjs, 228, 21

\bibitem[{{Clarke}(1945)}]{C45}
{Clarke}, A.~C. 1945, {Wireless World}, 55, 305

\bibitem[{{Delrez} {et~al.}(2018){Delrez}, {Gillon}, {Triaud}, {Demory}, {de
  Wit}, {Ingalls}, {Agol}, {Bolmont}, {Burdanov}, {Burgasser}, {Carey},
  {Jehin}, {Leconte}, {Lederer}, {Queloz}, {Selsis}, \& {Van Grootel}}]{DGT+18}
{Delrez}, L., {Gillon}, M., {Triaud}, A.~H.~M.~J., {et~al.} 2018, ArXiv
  e-prints, arXiv:1801.02554

\bibitem[{{Dyson}(1960)}]{D60}
{Dyson}, F.~J. 1960, Science, 131, 1667

\bibitem[{{Editorial}(2009)}]{SETI09}
{Editorial}. 2009, \nat, 461, 316

\bibitem[{{Enriquez} {et~al.}(2017){Enriquez}, {Siemion}, {Foster}, {Gajjar},
  {Hellbourg}, {Hickish}, {Isaacson}, {Price}, {Croft}, {DeBoer}, {Lebofsky},
  {MacMahon}, \& {Werthimer}}]{ESF+17}
{Enriquez}, J.~E., {Siemion}, A., {Foster}, G., {et~al.} 2017, \apj, 849, 104

\bibitem[{{Forgan} \& {Elvis}(2011)}]{FE11}
{Forgan}, D.~H., \& {Elvis}, M. 2011, International Journal of Astrobiology,
  10, 307

\bibitem[{{Grenfell}(2017)}]{G17}
{Grenfell}, J.~L. 2017, ArXiv e-prints, arXiv:1710.03976

\bibitem[{{Griffith} {et~al.}(2015){Griffith}, {Wright}, {Maldonado}, {Povich},
  {Sigur{\d}sson}, \& {Mullan}}]{GWM+15}
{Griffith}, R.~L., {Wright}, J.~T., {Maldonado}, J., {et~al.} 2015, \apjs, 217,
  25

\bibitem[{{Harp} {et~al.}(2016){Harp}, {Richards}, {Tarter}, {Dreher},
  {Jordan}, {Shostak}, {Smolek}, {Kilsdonk}, {Wilcox}, {Wimberly}, {Ross},
  {Barott}, {Ackermann}, \& {Blair}}]{HRT+16}
{Harp}, G.~R., {Richards}, J., {Tarter}, J.~C., {et~al.} 2016, \aj, 152, 181

\bibitem[{{Harris}(2002)}]{H02}
{Harris}, M.~J. 2002, Journal of the British Interplanetary Society, 55, 383

\bibitem[{{Hatchett} {et~al.}(2018){Hatchett}, {Barnes}, {Ahlers}, {MacKenzie},
  \& {Hedman}}]{HBA+18}
{Hatchett}, W.~T., {Barnes}, J.~W., {Ahlers}, J.~P., {MacKenzie}, S.~M., \&
  {Hedman}, M.~M. 2018, \na, 60, 88

\bibitem[{{Hecht}(2016)}]{H16}
{Hecht}, J. 2016, \nat, 530, 272

\bibitem[{{Hedman}(2015)}]{H15}
{Hedman}, M.~M. 2015, \apjl, 801, L33

\bibitem[{Hunter(2007)}]{H07}
Hunter, J.~D. 2007, Computing In Science \& Engineering, 9, 90

\bibitem[{{Kasting} {et~al.}(1993){Kasting}, {Whitmire}, \& {Reynolds}}]{KWR93}
{Kasting}, J.~F., {Whitmire}, D.~P., \& {Reynolds}, R.~T. 1993, \icarus, 101,
  108

\bibitem[{{Kawahara} \& {Fujii}(2010)}]{KJ10}
{Kawahara}, H., \& {Fujii}, Y. 2010, \apj, 720, 1333

\bibitem[{{Kopparapu}(2013)}]{K13}
{Kopparapu}, R.~K. 2013, \apjl, 767, L8

\bibitem[{{Korpela} {et~al.}(2015){Korpela}, {Sallmen}, \& {Leystra
  Greene}}]{KSLG15}
{Korpela}, E.~J., {Sallmen}, S.~M., \& {Leystra Greene}, D. 2015, \apj, 809,
  139

\bibitem[{{Lingam} \& {Loeb}(2017)}]{LL17a}
{Lingam}, M., \& {Loeb}, A. 2017, \apjl, 837, L23

\bibitem[{Mamajek {et~al.}(2012)Mamajek, Quillen, Pecaut, Moolekamp, Scott,
  Kenworthy, Cameron, \& Parley}]{MQP+12}
Mamajek, E.~E., Quillen, A.~C., Pecaut, M.~J., {et~al.} 2012, The Astronomical
  Journal, 143, 72.
\newblock \url{http://stacks.iop.org/1538-3881/143/i=3/a=72}

\bibitem[{P\'erez \& Granger(2007)}]{ipython07}
P\'erez, F., \& Granger, B.~E. 2007, Computing in Science and Engineering, 9,
  21.
\newblock \url{http://ipython.org}

\bibitem[{{Quanz}(2015)}]{Q15}
{Quanz}, S.~P. 2015, European Planetary Science Congress, 10, EPSC2015

\bibitem[{{Schlichting} \& {Chang}(2011)}]{SC17}
{Schlichting}, H.~E., \& {Chang}, P. 2011, \apj, 734, 117

\bibitem[{{Schneider} {et~al.}(2010){Schneider}, {L{\'e}ger}, {Fridlund},
  {White}, {Eiroa}, {Henning}, {Herbst}, {Lammer}, {Liseau}, {Paresce},
  {Penny}, {Quirrenbach}, {R{\"o}ttgering}, {Selsis}, {Beichman}, {Danchi},
  {Kaltenegger}, {Lunine}, {Stam}, \& {Tinetti}}]{SLF+10}
{Schneider}, J., {L{\'e}ger}, A., {Fridlund}, M., {et~al.} 2010, Astrobiology,
  10, 121

\bibitem[{{Schwieterman} {et~al.}(2017){Schwieterman}, {Kiang}, {Parenteau},
  {Harman}, {DasSarma}, {Fisher}, {Arney}, {Hartnett}, {Reinhard}, {Olson},
  {Meadows}, {Cockell}, {Walker}, {Grenfell}, {Hegde}, {Rugheimer}, {Hu}, \&
  {Lyons}}]{SKP+17}
{Schwieterman}, E.~W., {Kiang}, N.~Y., {Parenteau}, M.~N., {et~al.} 2017, ArXiv
  e-prints, arXiv:1705.05791

\bibitem[{{Seager} {et~al.}(2009){Seager}, {Deming}, \& {Valenti}}]{SDV09}
{Seager}, S., {Deming}, D., \& {Valenti}, J.~A. 2009, Astrophysics and Space
  Science Proceedings, 10, 123

\bibitem[{Van Der~Walt {et~al.}(2011)Van Der~Walt, Colbert, \&
  Varoquaux}]{numpy11}
Van Der~Walt, S., Colbert, S.~C., \& Varoquaux, G. 2011, Computing in Science
  \& Engineering, 13, 22

\bibitem[{{Van Grootel} {et~al.}(2017){Van Grootel}, {Fernandes}, {Gillon},
  {Jehin}, {Manfroid}, {Scuflaire}, {Burgasser}, {Burdanov}, {Delrez},
  {Demory}, {de Wit}, {Queloz}, \& {Triaud}}]{vGFG+17}
{Van Grootel}, V., {Fernandes}, C.~S., {Gillon}, M., {et~al.} 2017, ArXiv
  e-prints, arXiv:1712.01911

\bibitem[{{Wang} {et~al.}(2017){Wang}, {Wu}, {Barclay}, \& {Laughlin}}]{WWB+17}
{Wang}, S., {Wu}, D.-H., {Barclay}, T., \& {Laughlin}, G.~P. 2017, ArXiv
  e-prints, arXiv:1704.04290

\bibitem[{{Wright} {et~al.}(2016){Wright}, {Cartier}, {Zhao}, {Jontof-Hutter},
  \& {Ford}}]{WCZ+17}
{Wright}, J.~T., {Cartier}, K.~M.~S., {Zhao}, M., {Jontof-Hutter}, D., \&
  {Ford}, E.~B. 2016, \apj, 816, 17

\bibitem[{{Wright} {et~al.}(2014{\natexlab{a}}){Wright}, {Griffith},
  {Sigurdsson}, {Povich}, \& {Mullan}}]{WGS+14}
{Wright}, J.~T., {Griffith}, R.~L., {Sigurdsson}, S., {Povich}, M.~S., \&
  {Mullan}, B. 2014{\natexlab{a}}, \apj, 792, 27

\bibitem[{{Wright} {et~al.}(2014{\natexlab{b}}){Wright}, {Mullan},
  {Sigurdsson}, \& {Povich}}]{WMS+14}
{Wright}, J.~T., {Mullan}, B., {Sigurdsson}, S., \& {Povich}, M.~S.
  2014{\natexlab{b}}, \apj, 792, 26

\bibitem[{{Wright} {et~al.}(2014{\natexlab{c}}){Wright}, {Larkin}, {Moore},
  {Do}, {Simard}, {Adamkovics}, {Armus}, {Barth}, {Barton}, {Boyce}, {Cooke},
  {Cote}, {Davidge}, {Ellerbroek}, {Ghez}, {Liu}, {Lu}, {Macintosh}, {Mao},
  {Marois}, {Schoeck}, {Suzuki}, {Tan}, {Treu}, {Wang}, \& {Weiss}}]{WLM+14}
{Wright}, S.~A., {Larkin}, J.~E., {Moore}, A.~M., {et~al.} 2014{\natexlab{c}},
  in \procspie, Vol. 9147, Ground-based and Airborne Instrumentation for
  Astronomy V, 91479S

\end{thebibliography}

%\end{thebibliography}


  \acknowledgments The author gratefully acknowledges financial
  support from the Spanish Ministry of Economy and Competitivity
  through project AYA2014-60476-P. This research has made use of
  NASA's Astrophysics Data System Bibliographic Services.  The Python
  Matplotlib (\citealt{H07}), Astropy (\citealt{ACRT+13}), Scipy
  (\url{http://www.scipy.org}), Numpy (\citealt{numpy11}) and iPython
  (\citealt{ipython07}) modules have been used to generate the figures
  and some calculations for this paper. The author thanks the
  anonymous referee for valuable comments that helped improve an
  earlier manuscript.


\end{document}